\documentclass[sigconf]{acmart}
\AtBeginDocument{%
  \providecommand\BibTeX{{%
    \normalfont B\kern-0.5em{\scshape i\kern-0.25em b}\kern-0.8em\TeX}}}

\copyrightyear{2024}
\acmYear{2024}
\setcopyright{acmlicensed}\acmConference[GECCO '24 Companion]{Genetic and Evolutionary Computation Conference}{July 14--18, 2024}{Melbourne, VIC, Australia}
\acmBooktitle{Genetic and Evolutionary Computation Conference (GECCO '24 Companion), July 14--18, 2024, Melbourne, VIC, Australia}
\acmDOI{10.1145/3638530.3664121}
\acmISBN{979-8-4007-0495-6/24/07}

\usepackage{algorithm}
\usepackage{algorithmicx}
\usepackage{algpseudocode}
\usepackage{amsmath,amsfonts,bm}
\usepackage{pifont}
\usepackage{multirow}
\usepackage{booktabs}
\usepackage{colortbl}

\usepackage[most]{tcolorbox}

\definecolor{light-gray}{gray}{0.8}

\graphicspath{{./figs/}}

\usepackage{xspace}

\newcommand{\etal}{\textit{et al.}\xspace}

\newcommand{\figref}[1]{\figurename~\ref{#1}}
\newcommand{\secref}[1]{Section.~\ref{#1}}
\newcommand{\tabref}[1]{\tablename~\ref{#1}}

\newcommand{\bheading}[1]{{\vspace{0.4\baselineskip}\noindent{\textbf{#1}}}}
\newcommand{\lheading}[1]{{\vspace{0.4\baselineskip}\noindent{\textit{#1}}}}

\begin{document}

\title{L-AutoDA: Large Language Models for Automatically Evolving Decision-based Adversarial Attacks}

\author{Ping Guo}
\orcid{0000-0001-5412-914X}
\affiliation{%
    \institution{City University of Hong Kong}
    \streetaddress{Tat Chee Road 83}
    \city{Hong Kong}
    \country{Hong Kong SAR}
}
\affiliation{%
    \institution{CityU Shenzhen Research Institute}
    \city{Shenzhen}
    \country{China}
}
\email{pingguo5-c@my.cityu.edu.hk}

\author{Fei Liu}
\orcid{0000-0001-6719-0409}
\affiliation{%
    \institution{City University of Hong Kong}
    \streetaddress{Tat Chee Road 83}
    \city{Hong Kong}
    \country{Hong Kong SAR}
}
\affiliation{%
    \institution{CityU Shenzhen Research Institute}
    \city{Shenzhen}
    \country{China}
}
\email{fliu36-c@my.cityu.edu.hk}

\author{Xi Lin}
\orcid{0000-0001-5298-6893}
\affiliation{%
    \institution{City University of Hong Kong}
    \streetaddress{Tat Chee Road 83}
    \city{Hong Kong}
    \country{Hong Kong SAR}
}
\affiliation{%
    \institution{CityU Shenzhen Research Institute}
    \city{Shenzhen}
    \country{China}
}
\email{xi.lin@my.cityu.edu.hk}

\author{Qingchuan Zhao}

\orcid{0000-0003-0163-2846}
\affiliation{%
    \institution{City University of Hong Kong}
    \streetaddress{Tat Chee Road 83}
    \city{Hong Kong}
    \country{Hong Kong SAR}
}
\email{qizhao@cityu.edu.hk}

\author{Qingfu Zhang}

\orcid{0000-0003-0786-0671}
\affiliation{%
    \institution{City University of Hong Kong}
    \streetaddress{Tat Chee Road 83}
    \city{Hong Kong}
    \country{Hong Kong SAR}
}
\affiliation{%
    \institution{CityU Shenzhen Research Institute}
    \city{Shenzhen}
    \country{China}
}
\email{qingfu.zhang@cityu.edu.hk}


\begin{abstract}
    In the rapidly evolving field of machine learning, adversarial attacks pose a significant threat to the robustness and security of models. Amongst these, decision-based attacks are particularly insidious due to their nature of requiring only the model's decision output, which makes them notably challenging to counteract. This paper presents L-AutoDA (Large Language Model-based Automated Decision-based Adversarial Attacks), an innovative methodology that harnesses the generative capabilities of large language models (LLMs) to streamline the creation of such attacks. L-AutoDA employs an evolutionary strategy, where iterative interactions with LLMs lead to the autonomous generation of potent attack algorithms, thereby reducing human intervention. The performance of L-AutoDA was evaluated on the CIFAR-10 dataset, where it demonstrated substantial superiority over existing baseline methods in terms of success rate and computational efficiency. Ultimately, our results highlight the formidable utility of language models in crafting adversarial attacks and reveal promising directions for constructing more resilient AI systems.
\end{abstract}






\maketitle

\section{Introduction}

Deep neural network (DNN) models, despite their remarkable performance across a broad spectrum of domains, remain susceptible to \emph{adversarial attacks}~\cite{szegedy:2014:intriguing,goodfellow:2015:explaining}, which involve imperceptibly altering the input data to induce incorrect model responses. Such vulnerabilities threaten the integrity and reliability of machine learning applications, particularly in safety-critical scenarios such as autonomous vehicle driving~\cite{cao:2019:adversarial} and medical diagnostics~\cite{dong:2023:adversarial}. Attackers can engineer \emph{white-box attacks} using comprehensive knowledge of the DNN, or resort to \emph{black-box attacks} when the model's details are concealed~\cite{papernot:2017:practical}. Of particular concern are \emph{decision-based attacks} that necessitate only the model's output label information~\cite{ilyas:2018:blackbox}, posing a significant risk to real-world machine applications, such as commercial platforms that generally provide only the decision to users, thereby substantially endangering security and presenting challenges in implementing effective defenses~\cite{feng:2023:stateful,guo:2024:puridefense}.

The escalating arms race in trustworthy artificial intelligence (AI) domain, characterized by the rapid advancement of attack methodologies and the concurrent evolution of defensive strategies~\cite{Chakraborty:2021:survey,madry:2018:towards,wang:2024:adashield}, highlights the imperative for automating the generation and testing of adversarial attack algorithms~\cite{li:2023:arms}.
This necessity is particularly acute in the realm of decision-based attacks, which demand extensive manual labor to develop and refine strategic methodologies. Current approaches to decision-based attacks are heavily reliant on handcrafted heuristics~\cite{dong:2019:efficient,brendel:2018:decision,chen:2020:hopskipjumpattack,cheng:2020:signopt,cheng:2019:queryefficient}, posing significant impediments to enhancing their efficiency and efficacy.

The automation of adversarial attack algorithm design, underpinned by automatic program synthesis~\cite{gulwani2017program}, entails the generation of programs within complex constraints.
This area of research, known within the machine learning community as AutoML~\cite{feurer:2015:efficient}, seeks to devise strategies with minimal manual intervention.
AutoDA~\cite{fu:2022:autoda} represents the cutting-edge effort in this domain, adopting a random search across a thoughtfully assembled set of algebraic operations to engineer adversarial attack algorithms.
However, their method is inherently labor-intensive, particularly in developing domain-specific languages and establishing automated testing infrastructures.
Despite the intense investment of efforts and resources, the independent progression of novel algorithms without human expertise presents a substantial challenge~\cite{fu:2022:autoda,real:2020:automlzero}.

Recent literature has highlighted the potential of large language models (LLMs) for autonomous algorithm design, as demonstrated by initiatives such as Google's FunSearch~\cite{romera:2023:mathematical} and the evolutionary algorithm community's AEL~\cite{liu2023algorithm}.
These efforts have corroborated the feasibility of LLMs in the independent generation of algorithms.
The advantages of LLMs are manifold:
they can decode natural language inputs, obviating the need for domain-specific language encodings and thereby enabling the creation of innovative algorithms beyond the limitations of traditional encoding methods.
Furthermore, LLMs can be smoothly integrated into prevailing testing frameworks, requiring only slight modifications to existing test scripts since they can output program code directly, circumventing the need for decoding intermediate encodings.
A comparative analysis of this approach with conventional manual algorithm design and automatic program synthesis is delineated in \tabref{tab:comparison}.

\begin{table}[t]
    \centering
    \caption{Comparison of strengths and weaknesses of different algorithm design approaches.\label{tab:comparison}}
    \resizebox{0.47\textwidth}{!}{
        \begin{tabular}{lcccc}
            \toprule
            \multirow[c]{2}{*}{\textbf{Method}} & \multirow[c]{2}{*}{\textbf{Time}} & \multicolumn{2}{c}{\textbf{Expertise}} & \multirow[c]{2}{*}{\textbf{Refinement}}                              \\
            \cmidrule{3-4}
                                                &                                   & \textbf{Domain}                        & \textbf{Extra}                          &                            \\
            \midrule
            Manual                              & 1-2 Months                        & \ding{51}                              & \ding{55}                               & \ding{51}                  \\
            Automatic Synthesis                 & 1-2 Months                        & \ding{51}                              & \ding{51}                               & \ding{55}                  \\
            \textbf{L-AutoDA (Ours)}            & \textbf{1-2 Days}                 & \textcolor{red}{\ding{55}}             & \textcolor{red}{\ding{55}}              & \textcolor{red}{\ding{51}} \\
            \bottomrule
        \end{tabular}
    }
\end{table}

In this research, we exploit the AEL framework for developing decision-based adversarial attacks, introducing L-AutoDA, a cutting-edge automated framework tailored for crafting such attacks.
To the best of our knowledge, this work constitutes the first attempt to utilize LLMs in the development and autonomous evaluation of adversarial attack algorithms.
By integrating meticulously devised prompts and a population-based methodology within the AEL framework, as detailed in \cite{liu2023algorithm}, we succeed in deriving innovative strategies.
Remarkably, the genesis of all initial algorithms originated exclusively from LLMs and did not depend on established human-centric design principles.
This signifies a groundbreaking shift away from conventional approaches, featuring a new paradigm in the autonomous generation of adversarial attack algorithms.


Our contributions are as follows:

\begin{itemize}
    \item We introduce L-AutoDA, an innovative automated framework that incorporates LLMs to develop decision-based adversarial attack algorithms, marking a pioneering attempt to employ LLMs in this domain and setting the stage for new paradigms in the field.
    \item Our comparative analysis demonstrates the superiority of LLMs in crafting adversarial attack algorithms over existing methods. The benefits are threefold: \emph{1)} they enable algorithm generation through natural language interactions, thereby reducing the dependence on human expertise; \emph{2)} they exhibit a proficiency to generate more potent algorithms than those conceived by human experts, and \emph{3)} they display a capability to produce algorithms that can be integrated seamlessly with existing testing codes.
    \item The experimental evaluation and analysis highlight the generated algorithms’ robust performance, surpassing those that are manually designed. This furnishes new insights into the construction of decision-based adversarial attacks.
\end{itemize}

\section{Related Works}

\subsection{Decision-based Adversarial Attacks}

Decision-based adversarial attacks constitute the most challenging scenarios for attackers, given that the only information available about the target model is the output label.
Despite this obstacle, they pose a considerable threat to machine learning applications.
A pioneering study by Ilyas~\etal~\cite{ilyas:2018:blackbox} demonstrated the use of Natural Evolution Strategies (NES) to optimize a surrogate function with a limited number of queries to the model.
Subsequent advancements have focused on refining gradient estimation techniques.
For example, the OPT attack framework introduced by Cheng~\etal~\cite{cheng:2019:queryefficient} reformulates the primary optimization challenge.
More sophisticated methods, such as the Sign-OPT~\cite{cheng:2020:signopt}, emphasize the direction of the gradient rather than its magnitude, while the HopSkipJump attack~\cite{chen:2020:hopskipjumpattack} incorporates efficient gradient estimation and combine it with a binary search to closely track the decision boundary. The effectiveness of decision-based attacks is further supported by strategies based on random walks, such as the Boundary Attack~\cite{brendel:2018:decision} and Evolutionary Attack~\cite{dong:2019:efficient}.

\subsection{Automatically Devising Adversarial Attacks.}

The field of adversarial machine learning has increasingly focused on the automated development of attack algorithms~\cite{yin:2023:generalizable,fu:2022:autoda}.
The evolution of attack methods has progressed from basic gradient-based methods such as the Fast Gradient Sign Method (FGSM), which relies on actual gradient data~\cite{goodfellow:2015:explaining}, to more sophisticated iterative and optimization-based methods, such as decision-based attacks that require only output label data~\cite{dong:2019:efficient,brendel:2018:decision,chen:2020:hopskipjumpattack,cheng:2020:signopt,cheng:2019:queryefficient}.
Significant research efforts have been invested in the autonomous generation of attack methods utilizing genetic algorithms and evolutionary strategies.
To address the prohibitive inefficiency of exploring an unbounded function space of attack algorithms, researchers have introduced a domain-specific language (DSL) to constrain the complexity of functions, thereby achieving the notable efficiency improvements of AutoDA over traditional attack methods~\cite{fu:2022:autoda}.
However, creating these algorithms continues to be a labor-intensive process, requiring specialized knowledge to formulate a DSL, develop an associated code generator, and design an appropriate testing framework.

\subsection{LLMs for Algorithm Design.}

The recent surge in LLMs' capabilities, coupled with their access to extensive training datasets, has significantly enhanced their performance across various research domains~\cite{zhao2023survey,kasneci2023chatgpt}.
Notably, they excel in executing diverse tasks in a zero-shot fashion~\cite{min2021recent,tian2023chatgpt,lee2023benefits,nori2023capabilities,cheng2023exploring,jablonka2023gpt,blocklove2023chip,he2023chateda,yu2023gpt,zheng2023can,zhang2023automl,zhou2022large,gu2023systematic}.
Such progress opens avenues for LLMs to generate and manipulate complex algorithmic structures.

In extending their application, LLMs are now instrumental in the innovation of several algorithmic frameworks.
They have been effectively integrated as black-box components in the development of evolutionary algorithms, neural architectures, Monte Carlo Tree Search algorithms, solutions for graph-based combinatorial optimization, genetic programming, and open-ended challenges~\cite{zhao2023survey}.
While engaging with LLMs through prompts is common, it may result in suboptimal outcomes.
A fusion of large language models with evolutionary computation has emerged as a revolutionary advancement~\cite{wu2024evolutionary}, facilitating  the self-enhancement of algorithms~\cite{liu2023algorithm,liu2024evolution}, programming codes~\cite{Lehman2024}, and mathematical functions~\cite{romera2023mathematical} through autonomous, iterative refinement within an evolutionary setting.



\begin{figure*}[t]
    \centering
    \includegraphics[width=0.96\textwidth]{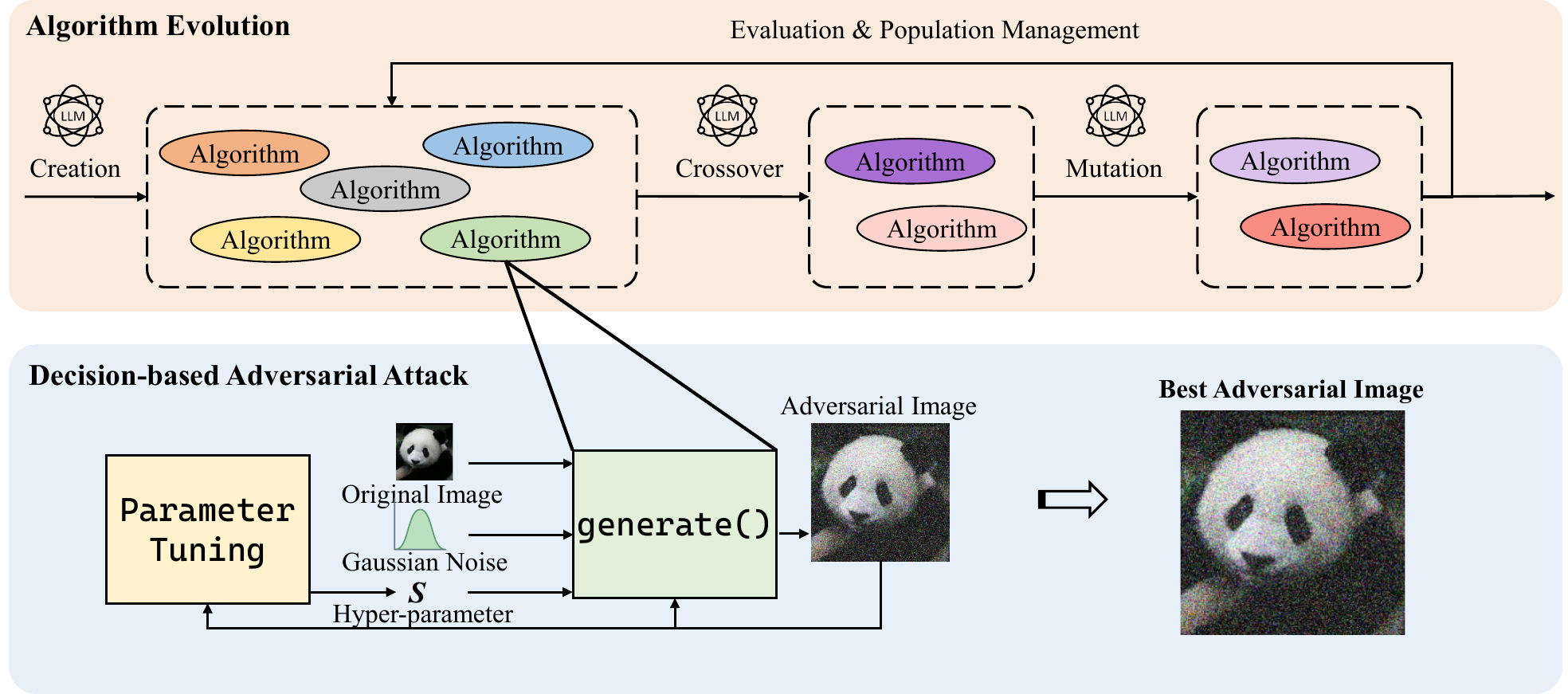}
    \caption{\textbf{Overview of the L-AutoDA Framework Methodology.} This diagram delineates the two core components of our L-AutoDA framework: the algorithm generation and testing phases. In the algorithm generation phase, we adopt the AEL framework, leveraging LLMs to guide an evolutionary search process. In the testing phase, we employ existing decision-based attack testing code, integrating these algorithms into the attack program to validate their efficacy.\label{fig:framework}}
    \Description{An overview of the algorithm framework.}
\end{figure*}

\section{Preliminaries}
\subsection{Decision-based Adversarial Attacks}
Consider a cloud-based image classifier $\mathcal{M}:\mathcal{X} \rightarrow \mathcal{Y}$, which maps images from an input space, denoted by $\mathcal{X}$, to an output space of classification probabilities, denoted by $\mathcal{Y}$.
The input space $\mathcal{X}$ consists of images with $C$ channels of $H \times W$ dimensions and is a subset of $[0,1]^{C \times H \times W}$, while the output space $\mathcal{Y}$, representing $m$ potential class labels, is a subset of $[0,1]^m$ probability space.

When attackers interact with this classifier, they submit a query image $\bm{x} \in \mathcal{X}$ and receive the predicted output $\mathcal{M}(\bm{x}) \in \mathcal{Y}$. The decision-making process can be clarified by defining the label of input $\bm{x}$ as $C(\bm{x}) = \arg\max_i\mathcal{M}_i(\bm{x})$, which indicates the model's highest confidence prediction. In decision-based attacks, the adversary only gains knowledge of this label.

The attacker's goal in a decision-based attack is to introduce a perturbation $\delta$ to the original input $\bm{x}_0$, where the perturbation is minimal yet effective such that $\|\delta\|_p \leq \epsilon$. This perturbation results in a modified input $\bm{x}_0+\delta$ that misleads the classifier into assigning a different label. Mathematically, this process is expressed as an optimization problem:
\begin{equation}
    \min \|\delta\|_p \quad \text{s.t.} \quad C(\bm{x}_0+\delta) \neq C(\bm{x}_0).
\end{equation}

To be deemed successful, a decision-based adversarial attack must ensure the perturbation's magnitude is within a predefined threshold, $\|\delta\|_p \leq \epsilon$.
While this work primarily examines \textit{untargeted} attacks bound by the $\ell_2$-norm ($p=2$), it also acknowledges the adaptability of our proposed method to facilitate \textit{targeted} attacks by altering the constraint to $C(\bm{x}_0+\delta) = y$, with $y$ being the designated target label.

\subsection{Algorithm Evolution using LLMs}

In this study, we employ the LLMs to generate the attacking heuristic in an evolutionary framework. It is structured around a cyclical process encompassing key EC stages including initialization, function evaluation, selection, crossover, mutation, and population management, the last of which meditates on diversity and convergence among the proposed solutions. Our methodology draws inspiration from the algorithmic approach developed by Liu~\etal~\cite{liu2023algorithm}.

\bheading{Initialization.}
The initial population is either derived from extant algorithms or is freshly generated using LLMs.
Using existing algorithms provides a solid baseline for the evolutionary search, whereas generation from scratch affords the possibility to discover an expansive, novel algorithmic domain.
Our methodology involves the latter, leveraging LLMs to generate an initial suite of algorithms. The exploration of evolution using established algorithms remains an integral aspect for formulating robust baselines in future research endeavors.

\bheading{Evaluating Algorithm.}
A pivotal component of AEL is assessing the solutions' fitness value.
We employ decision-based attack testing as the evaluation mechanism, defining fitness through the measurement of the $\ell_2$ distance between the original input and the adversarial output generated by the algorithm.

\bheading{Generating New Solutions.}
This stage adheres to the established protocols of EC.

\begin{itemize}
    \item \textbf{Selection.} Analogous to traditional EC practices, we select a predetermined number of algorithms to be retained through each iteration.
    \item \textbf{Crossover.} We facilitate the crossover operation by submitting a pair of algorithmic candidates, along with guiding prompts, to the LLMs, which in turn, synthesize a potentially superior algorithm. This approach leverages the LLMs' ability to boost the search process beyond the random search capabilities of automated program synthesis.
    \item \textbf{Mutation.} Introducing variation into the algorithmic pool is paramount for fostering diversity. This is accomplished by instructing the LLMs to introduce minor modifications to the current algorithms.
\end{itemize}

\section{L-AutoDA: LLM-based Automated Decision-based Adversarial Attacks}
In this section, we introduce our novel framework, L-AutoDA, which is designed for automatically generating decision-based adversarial attacks.
We begin by delineating the problem formulation and examining the search space associated with our framework (Section \ref{subsec:search}).
Subsequently, we describe the comprehensive structure of the L-AutoDA framework (Section \ref{subsec:framework}) as well as elaborate on the specifics of its implementation (Section \ref{subsec:implementation}).
An illustrative overview of the L-AutoDA architecture is depicted in \figref{fig:framework}.

\subsection{Decision-based Attack Framework\label{subsec:search}}
\bheading{Random Walk Template.}
We have developed a foundational framework for decision-based adversarial attacks, founded on the random walk paradigm, to establish the function search space, as depicted in Algorithm~\ref{alg:framework}.
This framework integrates critical elements from pioneering techniques such as the Evolutionary Attack~\cite{dong:2019:efficient}, the Boundary Attack~\cite{brendel:2018:decision}, and various other strategies.
Although gradient-based frameworks are also prevalent, we leave the exploration of this domain to future research endeavors.

The framework highlights two pivotal components for further improvement: the \texttt{generate} function and the accompanying hyperparameters.
The \texttt{generate} function is vital to the algorithm, handling the current adversarial sample $\bm{x}_1$, the original example $\bm{x}_0$, and synthesizing a new adversarial instance $\bm{x}$.
Hyperparameters are pivotal in steering the algorithm's behavior, influencing factors such as step size and the number of iterations. To streamline the search process, we adopted the parameter tuning strategy from \cite{fu:2022:autoda}, concentrating our efforts on refining the \texttt{generate} function.

\bheading{Search Space.}
We let  the LLM to explore the search space of the \texttt{generate} function.
While devising a comprehensive algorithm for the generation of perturbations is a viable approach to advance our LLM-based algorithmic framework, the extensive search space complicates the discovery of the optimum algorithmic solution.
Future investigations will engage with the wide-ranging possibilities and address the challenges arising from this extensive search space~\cite{gulwani2017program}.

\begin{algorithm}[t]
    \caption{Random Walk Framework for Decision-Based Attacks under $\ell_2$ perturbation\label{alg:framework}}
    \begin{algorithmic}[1]
        \State \textbf{Input:} original example $\bm{x}_0$, adversarial starting point $\bm{x}_1$
        \State \textbf{Output:} An adversarial example $\bm{x}$.
        \State \textbf{Initialization:} $\bm{x} \gets \bm{x}_1$; $d_{\min}\gets \|\bm{x}-\bm{x}_0\|_2$
        \While{query budget not reached}
        \State $\bm{x}'\gets \texttt{generate}(\bm{x},\bm{x}_0)$
        \If{$\bm{x}'$ is adversarial and $\|\bm{x}'-\bm{x}_0\|_2<d_{\min}$}
        \State $\bm{x}\gets \bm{x}'$; $d_{\min}\gets \|\bm{x}'-\bm{x}_0\|_2$
        \EndIf
        \State Update hyper-parameters.
        \EndWhile
        \State \textbf{return} $\bm{x}$
    \end{algorithmic}
\end{algorithm}

\subsection{L-AutoDA\label{subsec:framework}}
The L-AutoDA framework represents a cutting-edge system that leverages the AEL paradigm~\cite{liu2023algorithm} to expedite the creation of novel decision-based adversarial attack algorithms.
Central to L-AutoDA is the pursuit of an optimal \texttt{generate} function, which is responsible for generating new adversarial examples during the attack process.
The resulting \texttt{generate} functions are seamlessly integrated into existing decision-based attack programs, enhancing the continuous innovation and assessment of diverse attack strategies, as depicted in \figref{fig:framework}.
Subsequent paragraphs detail the workflow within the L-AutoDA framework, beginning with the initialization of a set of candidate algorithms.

\bheading{Initialization.}
As L-AutoDA adopts a population-based method to cultivate a diverse array of candidate algorithms, we first need to initialize the population. This process involves providing a carefully constructed prompt:

\begin{tcolorbox}
    [%
        enhanced,
        breakable,
        boxrule=0.1pt,
        left=1pt,
        right=1pt,
        bottom=1pt,
        top=1pt
    ]{
        \textbf{Initialization Prompt.}
        {\small Given an image \texttt{org\_img}, its adversarial image \texttt{best\_adv\_img}, and a random normal noise \texttt{std\_normal\_noise}, you need to design an algorithm to combine them to search for a new adversarial example \texttt{x\_new}. \texttt{hyperparams} ranges from 0.5 to 1.5.  It gets larger when this algorithm outputs more adversarial examples, and vice versa. It can be used to control the step size of the search. Operations you may use include: adding, subtracting, multiplying, dividing, dot product, and l2 norm computation. Design an novel algorithm with various search techniques. Your code should be able to run without further assistance.
        }
    }
\end{tcolorbox}

Moreover, the input and output parameters and their corresponding messages of the \texttt{generate} function are provided to the AEL framework to further ensure the legitimacy of the generated code.

\bheading{Population-based Search.}
Following initialization, L-AutoDA engages in a population-based search within the evolutionary computation paradigm, employing a specialized testing script (as mentioned in \secref{subsec:implementation}) to evaluate fitness values.

\lheading{Objective Value.}
The efficacy of the algorithms is measured by an objective value, denoted as the average distance between adversarial and original images. This value acts as a fitness function within the AEL framework and steers the evolutionary algorithm.

\lheading{Search Process}
Guided by the objective value, L-AutoDA applies evolutionary operations, such as selection, crossover, and mutation, to refine the assortment of algorithms.
Different from traditional evolutionary algorithms, L-AutoDA implements the above operations leveraging LLMs by interacting with them with prompts and information like the objective value.
During this process, the most promising candidates, or "elite" algorithms, are identified and retained. This evolutionary cycle is performed iteratively to enhance the development of more potent adversarial attack algorithms.

\bheading{Substantial Advantages.}
L-AutoDA's generative mechanism is harmoniously compatible with conventional decision-based attack programs. It assesses the quality of the generated algorithms by examining the output the attack program produces when provided with the generated \texttt{generate} function.
This methodology marks a significant leap from traditional program synthesis, which typically necessitates rigorous validation to confirm the legitimacy and functional integrity of the generated code.
By concentrating on the algorithmic output and its effectiveness, L-AutoDA streamlines the search process and demonstrates its superiority.

\subsection{Implementation}\label{subsec:implementation}
This section details the implementation of the L-AutoDA algorithm, starting with a comprehensive description of the search space for \texttt{generate} functions, followed by the elucidation of a feedback mechanism for hyperparameter adjustment. It concludes with an overview of the testing script used to evaluate the performance of the evolved algorithms.

\bheading{Function Specification.}
The \texttt{generate} function accepts four inputs: the original example $\bm{x}_0$, the adversarial starting point $\bm{x}_1$, standard random noise $\bm{r}$, and a dynamically adjusted hyperparameter $s$.  Its objective is to ingeniously integrate these inputs to produce an adversarial example $\bm{x}$, with the hyperparameter providing informed control over step size referencing.

\bheading{Hyper-parameter Tuning.}
Our approach to hyperparameter tuning adopts the strategy presented in \cite{fu:2022:autoda}. We introduce a piece-wise linear function $f(p)$ defined as:
\begin{equation}
    f(p) = \left\{\begin{matrix}
        0.5 + 2p                   & 0\leq p\leq 0.25 \\
        \frac{5}{6} + \frac{2p}{3} & 0.25 < p \leq 1  \\
    \end{matrix}\right.
\end{equation}
During each iteration, $p$ is updated in the following manner:
\begin{equation}
    p = 0.95p + 0.05k
\end{equation}
where $k$ represents the discovery of an improved adversarial point, taking on the value of $1$ if a better point is found and $0$ otherwise. The hyperparameter $s$ is then computed by:
\begin{equation}
    s = s \cdot [f(p)]^{0.1}
\end{equation}
This engenders a compensatory feedback loop, aimed at anchoring $p$ around 0.25.

\bheading{Testing Script.}
The AEL framework relies on a fitness function value to guide its evolutionary progress.
In this context, a testing script was devised to evaluate the efficacy of the algorithms produced.
To avoid the extensive time requirement associated with processing the entire test set, a representative subset of the dataset was chosen for our experiments.
These samples are used to compute the fitness value, utilizing standardized attack settings. This method employs a standardized set of attack parameters to calculate the fitness value. Although this approach may bring about a certain degree of bias, the empirical evaluation results support its effectiveness in accelerating the evolutionary search.

\section{Experiments}
\subsection{Experimental Setup}

\bheading{L-AutoDA Generation.}
The experimental setup for the L-AutoDA algorithm generation is divided into two distinct parts: \textit{1)} settings for the AEL running process and \textit{2)} for the objective value evaluation. Note that our experiments are conducted on CIFAR-10 dataset~\cite{CIFAR10} and a ResNet-18 classification model~\cite{resnet18_cifar}, which is a prevalent benchmark for adversarial attack algorithms.

\lheading{AEL Settings.}
In our setting, the AEL framework operates over 20 generations, each comprising 10 algorithm candidates. Moreover, we set the crossover probability at 1.0, ensuring that each pair of selected programs undergoes recombination, and the mutation probability at 0.5 to introduce variability. The default LLM for algorithm generation is \texttt{GPT-3.5-turbo-1106}, with plans to expand testing to additional large language models in subsequent research.

\lheading{Algorithm Evaluation.}
In assessing the performance of the devised algorithms, we have tailored our testing procedure to confine each algorithm to a maximum of 8,000 queries. We execute the algorithms on the first eight images of the CIFAR-10 test set to ensure a consistent and manageable testing environment. The adversarial images produced are then used to calculate the $\ell_2$ distances relative to their original counterparts. The mean of these distances is computed to serve as the fitness value, which is fed back into the AEL framework, thereby informing the evolutionary search for more effective attack algorithms.

\begin{table*}[t]
    \centering
    \caption{The full test performance of L-AutoDA-20 compared to three baseline algorithms.\label{tab:full_test}}
    \resizebox{0.72\textwidth}{!}{
        \begin{tabular}{lcccccc}
            \toprule
            Attack Name   & \multicolumn{3}{c}{Distance ($\ell_2$-norm)}              & \multicolumn{3}{c}{Attack Success Rate}                                                                                                                                                                                                       \\
            \cmidrule{2-4}\cmidrule{5-7}
            \# of Queries & $2500$                                                    & $5000$                                                    & $10000$                                                   & $2500$                                & $5000$                                & $10000$                               \\
            \midrule
            Boundary      & $1.9107_{1.2665}$                                         & $1.0938_{0.7861}$                                         & $0.4495_{0.3340}$                                         & \cellcolor{light-gray}$\mathbf{14.7}$ & \cellcolor{light-gray}$\mathbf{26.2}$ & $65.5$                                \\
            HSJA          & $2.0512_{1.0876}$                                         & $1.2833_{0.7442}$                                         & $0.8978_{0.5360}$                                         & $9.2$                                 & $16.1$                                & $24.6$                                \\
            HSJA*         & $2.6482_{1.5790}$                                         & $1.6532_{1.0347}$                                         & $1.1306_{0.6987}$                                         & $7.9$                                 & $13.9$                                & $19.6$                                \\
            L-AutoDA-20   & \cellcolor{light-gray}$\mathbf{1.5202}_{\mathbf{0.1337}}$ & \cellcolor{light-gray}$\mathbf{0.6171}_{\mathbf{0.1430}}$ & \cellcolor{light-gray}$\mathbf{0.3445}_{\mathbf{0.2386}}$ & $0.0$                                 & $0.5 $                                & \cellcolor{light-gray}$\mathbf{80.3}$ \\
            \bottomrule
        \end{tabular}
    }
\end{table*}

\begin{table}[t]
    \centering
    \caption{The performance of L-AutoDA compared to three baseline algorithms using the testing script. The mean distance of 1000 images are documented with the standard variance to be the subscript. The best performance cell is marked with light gray and the text within is bolded.\label{tab:performance}}
    \resizebox{0.33\textwidth}{!}{
        \begin{tabular}{cccc}
            \toprule
            \textbf{Boundary} & \textbf{HSJA} & \textbf{HSJA*} & \textbf{L-Auto-20}                      \\
            \midrule
            $0.3939$          & $1.3628$      & $1.2839$       & \cellcolor{light-gray}$\mathbf{0.2517}$ \\
            \bottomrule
        \end{tabular}
    }
\end{table}

\begin{figure}[t]
    \centering
    \includegraphics[width=0.41\textwidth]{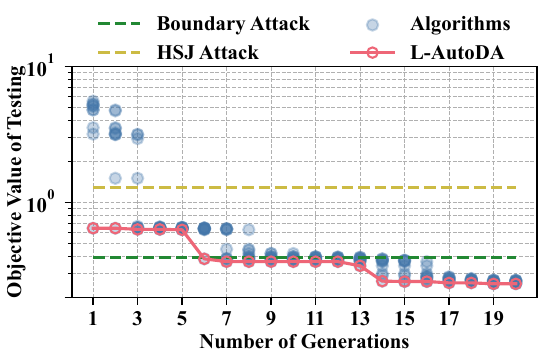}
    \caption{\textbf{Performance Trajectories of L-AutoDA.} This graph illustrates the comparative efficiency of our L-AutoDA framework against the human-best gradient-based (HopSkipJump Attack) and gradient-free (Boundary Attack) methods. L-AutoDA's candidates demonstrate a breakthrough in the 13th generation, surpassing the reference performance lines and continuing to enhance efficiency in subsequent generations. \label{fig:performance}}
\end{figure}

\bheading{Attack Evaluation.}
The evaluation process for different attacks is a crucial aspect of the experimental setup, providing a comprehensive assessment of the generated adversarial algorithms' performance.

\lheading{Datasets.}
Our evaluation utilizes a subset of the CIFAR-10 dataset, comprising 100 randomly sampled images from each class, to ensure a diverse and representative test bed. To facilitate a fair comparison across all attack algorithms, we introduce a set of 10 images with incorrect labels as the initial starting points for the attacks, ensuring that each algorithm begins from a standardized baseline.

\lheading{Comparative Algorithms.}
In our comparative analysis, we establish the Boundary attack~\cite{brendel:2018:decision}, which operates under the random walk framework, as the baseline algorithm. Additionally, we include the widely acknowledged SOTA decision-based attack algorithm, the HopSkipJumpAttack (HSJA)~\cite{chen:2020:hopskipjumpattack}, which employs a gradient-based approach. To further enrich our comparison, we introduce a variant of HSJA that utilizes a grid search strategy instead of its default geometric progression for step search, denoted as HSJA* in our paper. Our future work anticipates the inclusion of more attack algorithms for a more exhaustive comparison.

\lheading{Detailed Parameters.}
Delving into the detailed parameter settings, for the Boundary Attack, we set both the spherical and source steps for the Boundary Attack at 0.01, with a step size adaptation rate of 1.5. In the case of the HopSkipJump Attack, the parameter $\gamma$ is established at 1.0, initial gradient estimation starts with 100 steps, and is limited to a peak of 10,000 steps. Reflecting the adaptive nature of the L-AutoDA-generated algorithms, a negative feedback mechanism is employed to fine-tune the hyperparameter $s$, which is initially set to 0.001.

\subsection{Algorithm Generation}
The performance of the algorithms generated by L-AutoDA is encapsulated in \figref{fig:performance}, which demonstrates their compelling capabilities. Remarkably, the initial iteration of L-AutoDA produced algorithms that outperformed HSJA. Although this unexpected result may be partially attributed to the limited subset of images used during testing, it nonetheless underscores the potential of L-AutoDA in rapidly devising effective attack strategies. As the evolutionary process progressed, L-AutoDA continued to refine its algorithms, surpassing both HSJA and Boundary Attack by the 6th generation. This trend of improvement was consistent, with each subsequent generation enhancing the algorithms' effectiveness.

An intriguing aspect was the reduction in the variance of algorithm performance within each generation. This convergence suggests a stabilization of performance across the generated algorithms, indicating that L-AutoDA is not only producing more effective algorithms over time but also more reliable ones.

The results of the final round are documented in \tabref{tab:performance}. L-AutoDA's best algorithm within the 20th generation, denoted as L-Auto-20, achieved a mean perturbation distance of 0.2517 across the test images. This represents a significant improvement over the HSJA and Boundary Attack, which achieved mean perturbation distances of 1.3628 and 0.3939, respectively.

\subsection{Attack Evaluation}
To thoroughly evaluate the algorithms generated, we subjected them to tests on an expanded subset as delineated in our experimental setup. The most effective algorithm produced by the final iteration of L-AutoDA, referred to as L-AutoDA-20, was selected for benchmark comparison.

\bheading{Overall Results.}
We have documented the overall full test results in \tabref{tab:full_test}. The table reveals that L-AutoDA-20 is the most effective algorithm, achieving the lowest mean distance across all query counts. This result is particularly impressive given that L-AutoDA-20 was generated entirely from scratch by the LLM, without any human intervention. As for the success rate, L-AutoDA-20 achieved a 0\% success rate at 2500 queries, which is expected given the limited number of queries. The success rate then increased to 80.3\% at 10000 queries, surpassing all other algorithms. We delineate the relationship between attack success rate and distance in the following sections.

\begin{figure}[t]
    \centering
    \includegraphics[width=0.42\textwidth]{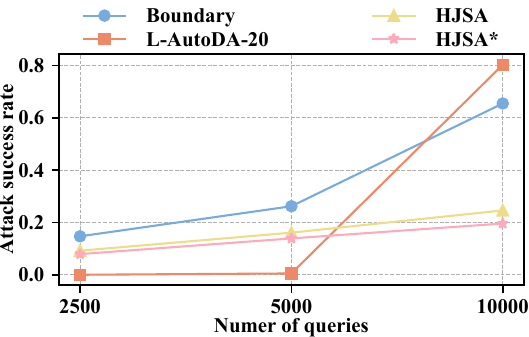}
    \caption{Attack Success Rate using different numbers of queries using L-AutoDA-20 and other attack algorithms.\label{fig:asr}}
\end{figure}

\bheading{Attack Success Rate.}
\figref{fig:asr} illustrates the attack success rate with the number of queries.  A successful attack is defined by an $\ell_2$ norm less than 0.5 between the adversarial example and the original image, consistent with the widely accepted standard in the current benchmarks~\cite{croce:2021:robustbench}. The figure reveals that L-AutoDA-20's performance is suboptimal at 2500 and 5000 queries. However, there is a notable uptick in success rate when the query count reaches 10000, surpassing all baseline algorithms. This pattern suggests that L-AutoDA sacrifices initial search efficiency to enhance the quality of the search at later stages, particularly after 8000 queries (testing script).

\begin{figure}[t]
    \centering
    \includegraphics[width=0.42\textwidth]{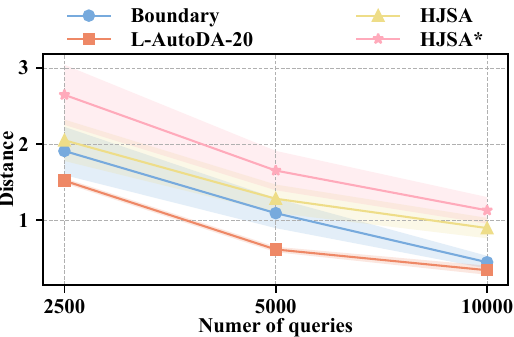}
    \caption{Distance between adversarial examples and original images using different numbers of queries using L-AutoDA-20 and other attack algorithms. The lines denote the mean value of the test pairs and the shaded areas represent a 0.25 multiplier of the standard deviation.\label{fig:distance}}
\end{figure}

\bheading{Distance.}
We present the comparative analysis of the perturbation distances in \figref{fig:distance}, where we plot the mean $\ell_2$ distance between the adversarial and original images against the number of queries used. The shaded areas in the figure represent a 0.25 multiplier of the standard deviation, providing insight into the variability of each algorithm's performance.

From \figref{fig:distance}, it is evident that L-AutoDA-20 maintains the most consistent performance across all tested query counts, as indicated by the smallest standard deviation values. This consistency suggests that L-AutoDA-20 is less sensitive to the variations in the input data, making it a robust choice for generating adversarial examples. Although this robustness may come at the cost of a reduced attack success rate in the initial phase, it becomes a significant advantage in later stages, particularly beyond 8000 queries.

The stability of L-AutoDA-20 is particularly beneficial when the attack requires subtlety, as it is capable of producing perturbations that are minimally perceptible yet still effective. This characteristic is crucial for scenarios where detectability is a concern and stealth is paramount.

\subsection{Additional Results on Median Distance}

To avoid the influence of variations with the images and better illustrate the effectiveness of our framework, we have demonstrated the median distance of the adversarial examples generated by different algorithms in \tabref{tab:median_distance}. The results are consistent with the previous analysis, with L-AutoDA-20 achieving the lowest median distance across all query counts.

\begin{table}[t]
    \centering
    \caption{Median distance of L-AutoDA-20 compared to three baseline algorithms. The best performance cell is marked with light gray and the text within is bolded.\label{tab:median_distance}}
    \resizebox{0.33\textwidth}{!}{
        \begin{tabular}{lccc}
            \toprule
                        & $2500$                                  & $5000$                                  & $10000$                                 \\
            \midrule
            Boundary    & $1.7374$                                & $0.9489$                                & $0.3695$                                \\
            HSJA        & $2.0230$                                & $1.2468$                                & $0.8646$                                \\
            HSJA*       & $2.5150$                                & $1.5580$                                & $1.0618$                                \\
            L-AutoDA-20 & \cellcolor{light-gray}$\mathbf{1.5301}$ & \cellcolor{light-gray}$\mathbf{0.5896}$ & \cellcolor{light-gray}$\mathbf{0.2862}$ \\
            \bottomrule
        \end{tabular}
    }
\end{table}

\subsection{Interpretation of the algorithms}

To elucidate the evolutionary process of the \texttt{generate()} function, a representative algorithm from the initial population and the most successful algorithm from the final population were chosen for comparative analysis.

\bheading{Initial Population.}
The selected algorithm from the initial population is detailed in Algorithm~\ref{alg:generate_init}.
While the search for adversarial examples is not assured by more efficient search vectors, this algorithm shows its flexibility by exploring different operations.
However, since we want to generate adversarial examples that are both effective and efficient, the initial algorithm may not be the most optimal choice and continue to evolve.

\bheading{Final Population.}
The \texttt{generate} function output by L-AutoDA is illustrated in Algorithm~\ref{alg:generate}. The algorithm starts by taking the difference between the original example $\mathbf{x}_0$ and the adversarial starting point $\mathbf{x}_1$. By moving along this vector, one can generate examples that are in between the original and the adversarial, which may help in exploring the space around known data points. Furthermore, efficient search is enabled through the inclusion of another normalized vector $\frac{\mathbf{d}}{norm}$. Then two scales of noise are added to the example, one with the same direction as the difference vector $\mathbf{d}$ and the other with the same direction as the normalized difference vector $\frac{\mathbf{d}}{norm}$. The noise is further scaled by a hyperparameter $s$ to control the magnitude of the perturbation. Combined with these search vectors, L-AutoDA is able to generate adversarial examples that are both effective and efficient.

\begin{algorithm}[t]
    \caption{\texttt{generate()} (Initial Population)\label{alg:generate_init}}
    \begin{algorithmic}[1]
        \State \textbf{Input:} original example $\mathbf{x}_0$, adversarial starting point $\mathbf{x}_1$,
        standard normal noise $\mathbf{n}$, hyperparameter $s$
        \State \textbf{Output:} A new proposed example $\mathbf{x}$
        \State $n_0 \gets \mathcal{N}(0,1)$
        \State $\mathbf{x} \gets s\mathbf{x}_0+(1-s)\mathbf{x}_1+n_0\mathbf{n}$
        \State $Operation = randChoice(add, sub, mul)$
        \State $n_1 \gets \mathcal{U}(0.5,1.5)$
        \If{$Operation == add$}
        \State $\mathbf{x} \gets \mathbf{x} + n_1\mathbf{n}$
        \ElsIf{$Operation == sub$}
        \State $\mathbf{x} \gets \mathbf{x} - n_1\mathbf{n}$
        \Else
        \State $\mathbf{x} \gets \mathbf{x} * n_1\mathbf{n}$
        \EndIf
        \State $\mathbf{d} \gets \mathbf{x}_0 - \mathbf{x}_1$
    \end{algorithmic}
\end{algorithm}
\begin{algorithm}[t]
    \caption{\texttt{generate()} (Final Population)\label{alg:generate}}
    \begin{algorithmic}[1]
        \State \textbf{Input:} original example $\mathbf{x}_0$, adversarial starting point $\mathbf{x}_1$,
        standard normal noise $\mathbf{n}$, hyperparameter $s$
        \State \textbf{Output:} A new proposed example $\mathbf{x}$
        \State $\mathbf{d} \gets \mathbf{x}_0 - \mathbf{x}_1$
        \State $norm = \max(\|\mathbf{d}\|_2, \|\mathbf{n}\|_2) $
        \State $\mathbf{x} \gets \mathbf{x}_1 + s(\mathbf{d}+\frac{\mathbf{d}}{norm}) + s(\mathbf{n} + s\frac{\mathbf{n}}{norm})$
    \end{algorithmic}
\end{algorithm}

\section{Discussion}

\bheading{Expanded Experimental Validation.}
Although our experimental framework, consisting of 20 generations with 10 individuals per generation, has yielded results surpassing those of manually-designed state-of-the-art algorithms, it has not fully tested the boundaries of our framework or LLMs. We will increase the number of generations and individuals to see if we can obtain better results. We aim to test these limits by increasing the population size and the number of generations. Additionally, initializing the search process with existing algorithms and subsequently refining them represents a promising avenue for further experimentation.

\bheading{Broader Algorithm Search Space.}
or expediency and as an initial attempt for automated attack algorithm design using LLMs into the automated design of attack algorithms using LLMs, we confined the search space to that defined by the \texttt{generate()} function. However, this narrow scope may restrict the discovery of optimal algorithms. Future work will seek to exploit the full potential of LLMs by allowing them to craft comprehensive algorithms without such constraints.

\bheading{Enhancing Prompt Adaptability.}
Our methodology employed a set of static prompts to assist LLMs in algorithm generation. However, the fixed prompts may not be the best prompts for LLMs to generate algorithms. The effectiveness of these prompts, however, may not represent an optimal use of LLM capabilities. The concept of chain-of-reasoning, which underpins our work and AEL, suggests a close relationship with adaptive prompt generation. Investigating methods of dynamically generating prompts is an objective of our ongoing research.

\bheading{Addressing Limitations.}
While the synthesis of programs using large language models is the focus of our research, it is not without its drawbacks. These models may occasionally yield unsatisfactory outcomes, albeit at a lower rate than traditional approaches. Improving the specificity of constraints within the prompts to ensure the validity of the algorithms produced will be an integral part of our forthcoming efforts.

\section{Conclusion}
In this paper, we have successfully demonstrated the innovative application of LLMs for the automatic design of decision-based adversarial attack algorithms. By leveraging the AEL framework, we have not only streamlined the algorithmic design process, but also achieved a significant reduction in the time and expertise required to develop effective adversarial attacks. Our approach, encapsulated in the L-AutoDA framework, represents a paradigm shift in the field of adversarial machine learning, showcasing the untapped potential of LLMs in the realm of security and algorithm synthesis.
\begin{acks}
    The work described in this paper was  supported by the Research Grants Council of the Hong Kong Special Administrative Region, China [GRF Project No. CityU 11215622], by Natural Science Foundation of China [Project No: 62276223] and by  Key Basic Research Foundation of Shenzhen, China.
\end{acks}

\bibliographystyle{ACM-Reference-Format}
\bibliography{000-paper}


\begin{thebibliography}{46}


\ifx \showCODEN    \undefined \def \showCODEN     #1{\unskip}     \fi
\ifx \showDOI      \undefined \def \showDOI       #1{#1}\fi
\ifx \showISBNx    \undefined \def \showISBNx     #1{\unskip}     \fi
\ifx \showISBNxiii \undefined \def \showISBNxiii  #1{\unskip}     \fi
\ifx \showISSN     \undefined \def \showISSN      #1{\unskip}     \fi
\ifx \showLCCN     \undefined \def \showLCCN      #1{\unskip}     \fi
\ifx \shownote     \undefined \def \shownote      #1{#1}          \fi
\ifx \showarticletitle \undefined \def \showarticletitle #1{#1}   \fi
\ifx \showURL      \undefined \def \showURL       {\relax}        \fi
\providecommand\bibfield[2]{#2}
\providecommand\bibinfo[2]{#2}
\providecommand\natexlab[1]{#1}
\providecommand\showeprint[2][]{arXiv:#2}

\bibitem[Blocklove et~al\mbox{.}(2023)]%
        {blocklove2023chip}
\bibfield{author}{\bibinfo{person}{Jason Blocklove}, \bibinfo{person}{Siddharth
  Garg}, \bibinfo{person}{Ramesh Karri}, {and} \bibinfo{person}{Hammond
  Pearce}.} \bibinfo{year}{2023}\natexlab{}.
\newblock \showarticletitle{Chip-Chat: Challenges and Opportunities in
  Conversational Hardware Design}.
\newblock \bibinfo{journal}{\emph{arXiv preprint arXiv:2305.13243}}
  (\bibinfo{year}{2023}).
\newblock


\bibitem[Brendel et~al\mbox{.}(2018)]%
        {brendel:2018:decision}
\bibfield{author}{\bibinfo{person}{Wieland Brendel}, \bibinfo{person}{Jonas
  Rauber}, {and} \bibinfo{person}{Matthias Bethge}.}
  \bibinfo{year}{2018}\natexlab{}.
\newblock \showarticletitle{Decision-Based Adversarial Attacks: Reliable
  Attacks Against Black-Box Machine Learning Models}. In
  \bibinfo{booktitle}{\emph{6th International Conference on Learning
  Representations, ({ICLR})}}. \bibinfo{publisher}{OpenReview.net}.
\newblock


\bibitem[Cao et~al\mbox{.}(2019)]%
        {cao:2019:adversarial}
\bibfield{author}{\bibinfo{person}{Yulong Cao}, \bibinfo{person}{Chaowei Xiao},
  \bibinfo{person}{Benjamin Cyr}, \bibinfo{person}{Yimeng Zhou},
  \bibinfo{person}{Won Park}, \bibinfo{person}{Sara Rampazzi},
  \bibinfo{person}{Qi~Alfred Chen}, \bibinfo{person}{Kevin Fu}, {and}
  \bibinfo{person}{Z.~Morley Mao}.} \bibinfo{year}{2019}\natexlab{}.
\newblock \showarticletitle{Adversarial Sensor Attack on LiDAR-based Perception
  in Autonomous Driving}. In \bibinfo{booktitle}{\emph{Proceedings of the 2019
  {ACM} {SIGSAC} Conference on Computer and Communications Security, {(CCS})}}.
  \bibinfo{publisher}{{ACM}}.
\newblock


\bibitem[Chakraborty et~al\mbox{.}(2021)]%
        {Chakraborty:2021:survey}
\bibfield{author}{\bibinfo{person}{Anirban Chakraborty},
  \bibinfo{person}{Manaar Alam}, \bibinfo{person}{Vishal Dey},
  \bibinfo{person}{Anupam Chattopadhyay}, {and} \bibinfo{person}{Debdeep
  Mukhopadhyay}.} \bibinfo{year}{2021}\natexlab{}.
\newblock \showarticletitle{A survey on adversarial attacks and defences}.
\newblock \bibinfo{journal}{\emph{{CAAI} Trans. Intell. Technol.}}
  (\bibinfo{year}{2021}).
\newblock
\urldef\tempurl%
\url{https://doi.org/10.1049/CIT2.12028}
\showDOI{\tempurl}


\bibitem[Chen et~al\mbox{.}(2020)]%
        {chen:2020:hopskipjumpattack}
\bibfield{author}{\bibinfo{person}{Jianbo Chen}, \bibinfo{person}{Michael~I.
  Jordan}, {and} \bibinfo{person}{Martin~J. Wainwright}.}
  \bibinfo{year}{2020}\natexlab{}.
\newblock \showarticletitle{HopSkipJumpAttack: {A} Query-Efficient
  Decision-Based Attack}. In \bibinfo{booktitle}{\emph{2020 {IEEE} Symposium on
  Security and Privacy, ({SP})}}. \bibinfo{publisher}{{IEEE}}.
\newblock


\bibitem[Cheng et~al\mbox{.}(2023)]%
        {cheng2023exploring}
\bibfield{author}{\bibinfo{person}{Kunming Cheng}, \bibinfo{person}{Qiang Guo},
  \bibinfo{person}{Yongbin He}, \bibinfo{person}{Yanqiu Lu},
  \bibinfo{person}{Shuqin Gu}, {and} \bibinfo{person}{Haiyang Wu}.}
  \bibinfo{year}{2023}\natexlab{}.
\newblock \showarticletitle{Exploring the potential of GPT-4 in biomedical
  engineering: the dawn of a new era}.
\newblock \bibinfo{journal}{\emph{Annals of Biomedical Engineering}}
  (\bibinfo{year}{2023}), \bibinfo{pages}{1--9}.
\newblock


\bibitem[Cheng et~al\mbox{.}(2019)]%
        {cheng:2019:queryefficient}
\bibfield{author}{\bibinfo{person}{Minhao Cheng}, \bibinfo{person}{Thong Le},
  \bibinfo{person}{Pin{-}Yu Chen}, \bibinfo{person}{Huan Zhang},
  \bibinfo{person}{Jinfeng Yi}, {and} \bibinfo{person}{Cho{-}Jui Hsieh}.}
  \bibinfo{year}{2019}\natexlab{}.
\newblock \showarticletitle{Query-Efficient Hard-label Black-box Attack: An
  Optimization-based Approach}. In \bibinfo{booktitle}{\emph{7th International
  Conference on Learning Representations, {ICLR}}}.
  \bibinfo{publisher}{OpenReview.net}.
\newblock


\bibitem[Cheng et~al\mbox{.}(2020)]%
        {cheng:2020:signopt}
\bibfield{author}{\bibinfo{person}{Minhao Cheng}, \bibinfo{person}{Simranjit
  Singh}, \bibinfo{person}{Patrick~H. Chen}, \bibinfo{person}{Pin{-}Yu Chen},
  \bibinfo{person}{Sijia Liu}, {and} \bibinfo{person}{Cho{-}Jui Hsieh}.}
  \bibinfo{year}{2020}\natexlab{}.
\newblock \showarticletitle{Sign-OPT: {A} Query-Efficient Hard-label
  Adversarial Attack}. In \bibinfo{booktitle}{\emph{8th International
  Conference on Learning Representations, {ICLR}}}.
  \bibinfo{publisher}{OpenReview.net}.
\newblock


\bibitem[Croce et~al\mbox{.}(2021)]%
        {croce:2021:robustbench}
\bibfield{author}{\bibinfo{person}{Francesco Croce}, \bibinfo{person}{Maksym
  Andriushchenko}, \bibinfo{person}{Vikash Sehwag}, \bibinfo{person}{Edoardo
  Debenedetti}, \bibinfo{person}{Nicolas Flammarion}, \bibinfo{person}{Mung
  Chiang}, \bibinfo{person}{Prateek Mittal}, {and} \bibinfo{person}{Matthias
  Hein}.} \bibinfo{year}{2021}\natexlab{}.
\newblock \showarticletitle{RobustBench: a standardized adversarial robustness
  benchmark}. In \bibinfo{booktitle}{\emph{Proceedings of the Neural
  Information Processing Systems Track on Datasets and Benchmarks 1,
  (NeurIPS)}}.
\newblock


\bibitem[Dadalto(2022)]%
        {resnet18_cifar}
\bibfield{author}{\bibinfo{person}{Eduardo Dadalto}.}
  \bibinfo{year}{2022}\natexlab{}.
\newblock \bibinfo{title}{ResNet18 trained on CIFAR10}.
\newblock
  \bibinfo{howpublished}{\url{https://huggingface.co/edadaltocg/resnet18_cifar10}}.
\newblock
\newblock
\shownote{Accessed: 2023-07-01}.


\bibitem[Dong et~al\mbox{.}(2023)]%
        {dong:2023:adversarial}
\bibfield{author}{\bibinfo{person}{Junhao Dong}, \bibinfo{person}{Junxi Chen},
  \bibinfo{person}{Xiaohua Xie}, \bibinfo{person}{Jianhuang Lai}, {and}
  \bibinfo{person}{Hao Chen}.} \bibinfo{year}{2023}\natexlab{}.
\newblock \showarticletitle{Adversarial Attack and Defense for Medical Image
  Analysis: Methods and Applications}.
\newblock \bibinfo{journal}{\emph{CoRR}} (\bibinfo{year}{2023}).
\newblock
\urldef\tempurl%
\url{https://doi.org/10.48550/ARXIV.2303.14133}
\showDOI{\tempurl}
\showeprint[arXiv]{2303.14133}


\bibitem[Dong et~al\mbox{.}(2019)]%
        {dong:2019:efficient}
\bibfield{author}{\bibinfo{person}{Yinpeng Dong}, \bibinfo{person}{Hang Su},
  \bibinfo{person}{Baoyuan Wu}, \bibinfo{person}{Zhifeng Li},
  \bibinfo{person}{Wei Liu}, \bibinfo{person}{Tong Zhang}, {and}
  \bibinfo{person}{Jun Zhu}.} \bibinfo{year}{2019}\natexlab{}.
\newblock \showarticletitle{Efficient Decision-Based Black-Box Adversarial
  Attacks on Face Recognition}. In \bibinfo{booktitle}{\emph{{IEEE} Conference
  on Computer Vision and Pattern Recognition, ({CVPR})}}.
  \bibinfo{publisher}{Computer Vision Foundation / {IEEE}}.
\newblock


\bibitem[Feng et~al\mbox{.}(2023)]%
        {feng:2023:stateful}
\bibfield{author}{\bibinfo{person}{Ryan Feng}, \bibinfo{person}{Ashish Hooda},
  \bibinfo{person}{Neal Mangaokar}, \bibinfo{person}{Kassem Fawaz},
  \bibinfo{person}{Somesh Jha}, {and} \bibinfo{person}{Atul Prakash}.}
  \bibinfo{year}{2023}\natexlab{}.
\newblock \showarticletitle{Stateful Defenses for Machine Learning Models Are
  Not Yet Secure Against Black-box Attacks}. In
  \bibinfo{booktitle}{\emph{Proceedings of the 2023 {ACM} {SIGSAC} Conference
  on Computer and Communications Security, ({CCS})}}.
  \bibinfo{publisher}{{ACM}}, \bibinfo{pages}{786--800}.
\newblock


\bibitem[Feurer et~al\mbox{.}(2015)]%
        {feurer:2015:efficient}
\bibfield{author}{\bibinfo{person}{Matthias Feurer}, \bibinfo{person}{Aaron
  Klein}, \bibinfo{person}{Katharina Eggensperger},
  \bibinfo{person}{Jost~Tobias Springenberg}, \bibinfo{person}{Manuel Blum},
  {and} \bibinfo{person}{Frank Hutter}.} \bibinfo{year}{2015}\natexlab{}.
\newblock \showarticletitle{Efficient and Robust Automated Machine Learning}.
  In \bibinfo{booktitle}{\emph{Advances in Neural Information Processing
  Systems 28: Annual Conference on Neural Information Processing Systems
  2015}}.
\newblock


\bibitem[Fu et~al\mbox{.}(2022)]%
        {fu:2022:autoda}
\bibfield{author}{\bibinfo{person}{Qi{-}An Fu}, \bibinfo{person}{Yinpeng Dong},
  \bibinfo{person}{Hang Su}, \bibinfo{person}{Jun Zhu}, {and}
  \bibinfo{person}{Chao Zhang}.} \bibinfo{year}{2022}\natexlab{}.
\newblock \showarticletitle{AutoDA: Automated Decision-based Iterative
  Adversarial Attacks}. In \bibinfo{booktitle}{\emph{31st {USENIX} Security
  Symposium, {USENIX} Security 2022}}. \bibinfo{publisher}{{USENIX}
  Association}, \bibinfo{pages}{3557--3574}.
\newblock


\bibitem[Goodfellow et~al\mbox{.}(2015)]%
        {goodfellow:2015:explaining}
\bibfield{author}{\bibinfo{person}{Ian~J. Goodfellow},
  \bibinfo{person}{Jonathon Shlens}, {and} \bibinfo{person}{Christian
  Szegedy}.} \bibinfo{year}{2015}\natexlab{}.
\newblock \showarticletitle{Explaining and Harnessing Adversarial Examples}. In
  \bibinfo{booktitle}{\emph{3rd International Conference on Learning
  Representations, ({ICLR})}}.
\newblock


\bibitem[Gu et~al\mbox{.}(2023)]%
        {gu2023systematic}
\bibfield{author}{\bibinfo{person}{Jindong Gu}, \bibinfo{person}{Zhen Han},
  \bibinfo{person}{Shuo Chen}, \bibinfo{person}{Ahmad Beirami},
  \bibinfo{person}{Bailan He}, \bibinfo{person}{Gengyuan Zhang},
  \bibinfo{person}{Ruotong Liao}, \bibinfo{person}{Yao Qin},
  \bibinfo{person}{Volker Tresp}, {and} \bibinfo{person}{Philip Torr}.}
  \bibinfo{year}{2023}\natexlab{}.
\newblock \showarticletitle{A systematic survey of prompt engineering on
  vision-language foundation models}.
\newblock \bibinfo{journal}{\emph{arXiv preprint arXiv:2307.12980}}
  (\bibinfo{year}{2023}).
\newblock


\bibitem[Gulwani et~al\mbox{.}(2017)]%
        {gulwani2017program}
\bibfield{author}{\bibinfo{person}{Sumit Gulwani}, \bibinfo{person}{Oleksandr
  Polozov}, \bibinfo{person}{Rishabh Singh}, {et~al\mbox{.}}}
  \bibinfo{year}{2017}\natexlab{}.
\newblock \showarticletitle{Program synthesis}.
\newblock \bibinfo{journal}{\emph{Foundations and Trends{\textregistered} in
  Programming Languages}} (\bibinfo{year}{2017}).
\newblock


\bibitem[Guo et~al\mbox{.}(2024)]%
        {guo:2024:puridefense}
\bibfield{author}{\bibinfo{person}{Ping Guo}, \bibinfo{person}{Zhiyuan Yang},
  \bibinfo{person}{Xi Lin}, \bibinfo{person}{Qingchuan Zhao}, {and}
  \bibinfo{person}{Qingfu Zhang}.} \bibinfo{year}{2024}\natexlab{}.
\newblock \showarticletitle{PuriDefense: Randomized Local Implicit Adversarial
  Purification for Defending Black-box Query-based Attacks}.
\newblock \bibinfo{journal}{\emph{arXiv preprint arXiv:2401.10586}}
  (\bibinfo{year}{2024}).
\newblock


\bibitem[He et~al\mbox{.}(2023)]%
        {he2023chateda}
\bibfield{author}{\bibinfo{person}{Zhuolun He}, \bibinfo{person}{Haoyuan Wu},
  \bibinfo{person}{Xinyun Zhang}, \bibinfo{person}{Xufeng Yao},
  \bibinfo{person}{Su Zheng}, \bibinfo{person}{Haisheng Zheng}, {and}
  \bibinfo{person}{Bei Yu}.} \bibinfo{year}{2023}\natexlab{}.
\newblock \showarticletitle{ChatEDA: A Large Language Model Powered Autonomous
  Agent for EDA}.
\newblock \bibinfo{journal}{\emph{arXiv preprint arXiv:2308.10204}}
  (\bibinfo{year}{2023}).
\newblock


\bibitem[Ilyas et~al\mbox{.}(2018)]%
        {ilyas:2018:blackbox}
\bibfield{author}{\bibinfo{person}{Andrew Ilyas}, \bibinfo{person}{Logan
  Engstrom}, \bibinfo{person}{Anish Athalye}, {and} \bibinfo{person}{Jessy
  Lin}.} \bibinfo{year}{2018}\natexlab{}.
\newblock \showarticletitle{Black-box Adversarial Attacks with Limited Queries
  and Information}. In \bibinfo{booktitle}{\emph{Proceedings of the 35th
  International Conference on Machine Learning, ({ICML})}}
  \emph{(\bibinfo{series}{Proceedings of Machine Learning Research})}.
  \bibinfo{publisher}{{PMLR}}.
\newblock


\bibitem[Jablonka et~al\mbox{.}(2023)]%
        {jablonka2023gpt}
\bibfield{author}{\bibinfo{person}{Kevin~Maik Jablonka},
  \bibinfo{person}{Philippe Schwaller}, \bibinfo{person}{Andres
  Ortega-Guerrero}, {and} \bibinfo{person}{Berend Smit}.}
  \bibinfo{year}{2023}\natexlab{}.
\newblock \showarticletitle{Is GPT-3 all you need for low-data discovery in
  chemistry?}
\newblock  (\bibinfo{year}{2023}).
\newblock


\bibitem[Kasneci et~al\mbox{.}(2023)]%
        {kasneci2023chatgpt}
\bibfield{author}{\bibinfo{person}{Enkelejda Kasneci}, \bibinfo{person}{Kathrin
  Se{\ss}ler}, \bibinfo{person}{Stefan K{\"u}chemann}, \bibinfo{person}{Maria
  Bannert}, \bibinfo{person}{Daryna Dementieva}, \bibinfo{person}{Frank
  Fischer}, \bibinfo{person}{Urs Gasser}, \bibinfo{person}{Georg Groh},
  \bibinfo{person}{Stephan G{\"u}nnemann}, \bibinfo{person}{Eyke
  H{\"u}llermeier}, {et~al\mbox{.}}} \bibinfo{year}{2023}\natexlab{}.
\newblock \showarticletitle{ChatGPT for good? On opportunities and challenges
  of large language models for education}.
\newblock \bibinfo{journal}{\emph{Learning and individual differences}}
  \bibinfo{volume}{103} (\bibinfo{year}{2023}), \bibinfo{pages}{102274}.
\newblock


\bibitem[Krizhevsky(2009)]%
        {CIFAR10}
\bibfield{author}{\bibinfo{person}{A. Krizhevsky}.}
  \bibinfo{year}{2009}\natexlab{}.
\newblock \bibinfo{booktitle}{\emph{{Learning Multiple Layers of Features from
  Tiny Images}}}.
\newblock \bibinfo{type}{Technical Report}. \bibinfo{institution}{Univ.
  Toronto}.
\newblock


\bibitem[Lee et~al\mbox{.}(2023)]%
        {lee2023benefits}
\bibfield{author}{\bibinfo{person}{Peter Lee}, \bibinfo{person}{Sebastien
  Bubeck}, {and} \bibinfo{person}{Joseph Petro}.}
  \bibinfo{year}{2023}\natexlab{}.
\newblock \showarticletitle{Benefits, limits, and risks of GPT-4 as an AI
  chatbot for medicine}.
\newblock \bibinfo{journal}{\emph{New England Journal of Medicine}}
  \bibinfo{volume}{388}, \bibinfo{number}{13} (\bibinfo{year}{2023}),
  \bibinfo{pages}{1233--1239}.
\newblock


\bibitem[Lehman et~al\mbox{.}(2024)]%
        {Lehman2024}
\bibfield{author}{\bibinfo{person}{Joel Lehman}, \bibinfo{person}{Jonathan
  Gordon}, \bibinfo{person}{Shawn Jain}, \bibinfo{person}{Kamal Ndousse},
  \bibinfo{person}{Cathy Yeh}, {and} \bibinfo{person}{Kenneth~O. Stanley}.}
  \bibinfo{year}{2024}\natexlab{}.
\newblock \bibinfo{booktitle}{\emph{Evolution Through Large Models}}.
\newblock \bibinfo{publisher}{Springer Nature Singapore},
  \bibinfo{address}{Singapore}, \bibinfo{pages}{331--366}.
\newblock


\bibitem[Li et~al\mbox{.}(2023)]%
        {li:2023:arms}
\bibfield{author}{\bibinfo{person}{Deqiang Li}, \bibinfo{person}{Qianmu Li},
  \bibinfo{person}{Yanfang~(Fanny) Ye}, {and} \bibinfo{person}{Shouhuai Xu}.}
  \bibinfo{year}{2023}\natexlab{}.
\newblock \showarticletitle{Arms Race in Adversarial Malware Detection: {A}
  Survey}.
\newblock \bibinfo{journal}{\emph{{ACM} Comput. Surv.}} (\bibinfo{year}{2023}).
\newblock


\bibitem[Liu et~al\mbox{.}(2024)]%
        {liu2024evolution}
\bibfield{author}{\bibinfo{person}{Fei Liu}, \bibinfo{person}{Xialiang Tong},
  \bibinfo{person}{Mingxuan Yuan}, \bibinfo{person}{Xi Lin},
  \bibinfo{person}{Fu Luo}, \bibinfo{person}{Zhenkun Wang},
  \bibinfo{person}{Zhichao Lu}, {and} \bibinfo{person}{Qingfu Zhang}.}
  \bibinfo{year}{2024}\natexlab{}.
\newblock \showarticletitle{Evolution of Heuristics: Towards Efficient
  Automatic Algorithm Design Using Large Language Mode}.
\newblock  (\bibinfo{year}{2024}).
\newblock
\showeprint[arXiv]{2401.02051}


\bibitem[Liu et~al\mbox{.}(2023)]%
        {liu2023algorithm}
\bibfield{author}{\bibinfo{person}{Fei Liu}, \bibinfo{person}{Xialiang Tong},
  \bibinfo{person}{Mingxuan Yuan}, {and} \bibinfo{person}{Qingfu Zhang}.}
  \bibinfo{year}{2023}\natexlab{}.
\newblock \showarticletitle{Algorithm Evolution Using Large Language Model}.
\newblock \bibinfo{journal}{\emph{arXiv preprint arXiv:2311.15249}}
  (\bibinfo{year}{2023}).
\newblock


\bibitem[Madry et~al\mbox{.}(2018)]%
        {madry:2018:towards}
\bibfield{author}{\bibinfo{person}{Aleksander Madry},
  \bibinfo{person}{Aleksandar Makelov}, \bibinfo{person}{Ludwig Schmidt},
  \bibinfo{person}{Dimitris Tsipras}, {and} \bibinfo{person}{Adrian Vladu}.}
  \bibinfo{year}{2018}\natexlab{}.
\newblock \showarticletitle{Towards Deep Learning Models Resistant to
  Adversarial Attacks}. In \bibinfo{booktitle}{\emph{6th International
  Conference on Learning Representations, ({ICLR})}}.
  \bibinfo{publisher}{OpenReview.net}.
\newblock


\bibitem[Min et~al\mbox{.}(2021)]%
        {min2021recent}
\bibfield{author}{\bibinfo{person}{Bonan Min}, \bibinfo{person}{Hayley Ross},
  \bibinfo{person}{Elior Sulem}, \bibinfo{person}{Amir Pouran~Ben Veyseh},
  \bibinfo{person}{Thien~Huu Nguyen}, \bibinfo{person}{Oscar Sainz},
  \bibinfo{person}{Eneko Agirre}, \bibinfo{person}{Ilana Heintz}, {and}
  \bibinfo{person}{Dan Roth}.} \bibinfo{year}{2021}\natexlab{}.
\newblock \showarticletitle{Recent advances in natural language processing via
  large pre-trained language models: A survey}.
\newblock \bibinfo{journal}{\emph{Comput. Surveys}} (\bibinfo{year}{2021}).
\newblock


\bibitem[Nori et~al\mbox{.}(2023)]%
        {nori2023capabilities}
\bibfield{author}{\bibinfo{person}{Harsha Nori}, \bibinfo{person}{Nicholas
  King}, \bibinfo{person}{Scott~Mayer McKinney}, \bibinfo{person}{Dean
  Carignan}, {and} \bibinfo{person}{Eric Horvitz}.}
  \bibinfo{year}{2023}\natexlab{}.
\newblock \showarticletitle{Capabilities of gpt-4 on medical challenge
  problems}.
\newblock \bibinfo{journal}{\emph{arXiv preprint arXiv:2303.13375}}
  (\bibinfo{year}{2023}).
\newblock


\bibitem[Papernot et~al\mbox{.}(2017)]%
        {papernot:2017:practical}
\bibfield{author}{\bibinfo{person}{Nicolas Papernot},
  \bibinfo{person}{Patrick~D. McDaniel}, \bibinfo{person}{Ian~J. Goodfellow},
  \bibinfo{person}{Somesh Jha}, \bibinfo{person}{Z.~Berkay Celik}, {and}
  \bibinfo{person}{Ananthram Swami}.} \bibinfo{year}{2017}\natexlab{}.
\newblock \showarticletitle{Practical Black-Box Attacks against Machine
  Learning}. In \bibinfo{booktitle}{\emph{Proceedings of the 2017 {ACM} on Asia
  Conference on Computer and Communications Security, {AsiaCCS}}}.
  \bibinfo{publisher}{{ACM}}.
\newblock


\bibitem[Real et~al\mbox{.}(2020)]%
        {real:2020:automlzero}
\bibfield{author}{\bibinfo{person}{Esteban Real}, \bibinfo{person}{Chen Liang},
  \bibinfo{person}{David~R. So}, {and} \bibinfo{person}{Quoc~V. Le}.}
  \bibinfo{year}{2020}\natexlab{}.
\newblock \showarticletitle{AutoML-Zero: Evolving Machine Learning Algorithms
  From Scratch}. In \bibinfo{booktitle}{\emph{Proceedings of the 37th
  International Conference on Machine Learning, ({ICML})}}
  \emph{(\bibinfo{series}{Proceedings of Machine Learning Research})}.
  \bibinfo{publisher}{{PMLR}}.
\newblock


\bibitem[Romera-Paredes et~al\mbox{.}(2023a)]%
        {romera:2023:mathematical}
\bibfield{author}{\bibinfo{person}{Bernardino Romera-Paredes},
  \bibinfo{person}{Mohammadamin Barekatain}, \bibinfo{person}{Alexander
  Novikov}, \bibinfo{person}{Matej Balog}, \bibinfo{person}{M~Pawan Kumar},
  \bibinfo{person}{Emilien Dupont}, \bibinfo{person}{Francisco~JR Ruiz},
  \bibinfo{person}{Jordan~S Ellenberg}, \bibinfo{person}{Pengming Wang},
  \bibinfo{person}{Omar Fawzi}, {et~al\mbox{.}}}
  \bibinfo{year}{2023}\natexlab{a}.
\newblock \showarticletitle{Mathematical discoveries from program search with
  large language models}.
\newblock \bibinfo{journal}{\emph{Nature}} (\bibinfo{year}{2023}),
  \bibinfo{pages}{1--3}.
\newblock


\bibitem[Romera-Paredes et~al\mbox{.}(2023b)]%
        {romera2023mathematical}
\bibfield{author}{\bibinfo{person}{Bernardino Romera-Paredes},
  \bibinfo{person}{Mohammadamin Barekatain}, \bibinfo{person}{Alexander
  Novikov}, \bibinfo{person}{Matej Balog}, \bibinfo{person}{M~Pawan Kumar},
  \bibinfo{person}{Emilien Dupont}, \bibinfo{person}{Francisco~JR Ruiz},
  \bibinfo{person}{Jordan~S Ellenberg}, \bibinfo{person}{Pengming Wang},
  \bibinfo{person}{Omar Fawzi}, {et~al\mbox{.}}}
  \bibinfo{year}{2023}\natexlab{b}.
\newblock \showarticletitle{Mathematical discoveries from program search with
  large language models}.
\newblock \bibinfo{journal}{\emph{Nature}} (\bibinfo{year}{2023}),
  \bibinfo{pages}{1--3}.
\newblock


\bibitem[Szegedy et~al\mbox{.}(2014)]%
        {szegedy:2014:intriguing}
\bibfield{author}{\bibinfo{person}{Christian Szegedy},
  \bibinfo{person}{Wojciech Zaremba}, \bibinfo{person}{Ilya Sutskever},
  \bibinfo{person}{Joan Bruna}, \bibinfo{person}{Dumitru Erhan},
  \bibinfo{person}{Ian~J. Goodfellow}, {and} \bibinfo{person}{Rob Fergus}.}
  \bibinfo{year}{2014}\natexlab{}.
\newblock \showarticletitle{Intriguing properties of neural networks}. In
  \bibinfo{booktitle}{\emph{2nd International Conference on Learning
  Representations ({ICLR})}}.
\newblock


\bibitem[Tian et~al\mbox{.}(2023)]%
        {tian2023chatgpt}
\bibfield{author}{\bibinfo{person}{Haoye Tian}, \bibinfo{person}{Weiqi Lu},
  \bibinfo{person}{Tsz~On Li}, \bibinfo{person}{Xunzhu Tang},
  \bibinfo{person}{Shing-Chi Cheung}, \bibinfo{person}{Jacques Klein}, {and}
  \bibinfo{person}{Tegawend{\'e}~F Bissyand{\'e}}.}
  \bibinfo{year}{2023}\natexlab{}.
\newblock \showarticletitle{Is ChatGPT the Ultimate Programming Assistant--How
  far is it?}
\newblock \bibinfo{journal}{\emph{arXiv preprint arXiv:2304.11938}}
  (\bibinfo{year}{2023}).
\newblock


\bibitem[Wang et~al\mbox{.}(2024)]%
        {wang:2024:adashield}
\bibfield{author}{\bibinfo{person}{Yu Wang}, \bibinfo{person}{Xiaogeng Liu},
  \bibinfo{person}{Yu Li}, \bibinfo{person}{Muhao Chen}, {and}
  \bibinfo{person}{Chaowei Xiao}.} \bibinfo{year}{2024}\natexlab{}.
\newblock \showarticletitle{AdaShield: Safeguarding Multimodal Large Language
  Models from Structure-based Attack via Adaptive Shield Prompting}.
\newblock \bibinfo{journal}{\emph{CoRR}} (\bibinfo{year}{2024}).
\newblock


\bibitem[Wu et~al\mbox{.}(2024)]%
        {wu2024evolutionary}
\bibfield{author}{\bibinfo{person}{Xingyu Wu}, \bibinfo{person}{Sheng-hao Wu},
  \bibinfo{person}{Jibin Wu}, \bibinfo{person}{Liang Feng}, {and}
  \bibinfo{person}{Kay~Chen Tan}.} \bibinfo{year}{2024}\natexlab{}.
\newblock \showarticletitle{Evolutionary Computation in the Era of Large
  Language Model: Survey and Roadmap}.
\newblock \bibinfo{journal}{\emph{arXiv preprint arXiv:2401.10034}}
  (\bibinfo{year}{2024}).
\newblock


\bibitem[Yin et~al\mbox{.}(2023)]%
        {yin:2023:generalizable}
\bibfield{author}{\bibinfo{person}{Fei Yin}, \bibinfo{person}{Yong Zhang},
  \bibinfo{person}{Baoyuan Wu}, \bibinfo{person}{Yan Feng},
  \bibinfo{person}{Jingyi Zhang}, \bibinfo{person}{Yanbo Fan}, {and}
  \bibinfo{person}{Yujiu Yang}.} \bibinfo{year}{2023}\natexlab{}.
\newblock \showarticletitle{Generalizable Black-Box Adversarial Attack with
  Meta Learning}.
\newblock \bibinfo{journal}{\emph{CoRR}}  \bibinfo{volume}{abs/2301.00364}
  (\bibinfo{year}{2023}).
\newblock
\showeprint[arXiv]{2301.00364}


\bibitem[Yu et~al\mbox{.}(2023)]%
        {yu2023gpt}
\bibfield{author}{\bibinfo{person}{Caiyang Yu}, \bibinfo{person}{Xianggen Liu},
  \bibinfo{person}{Chenwei Tang}, \bibinfo{person}{Wentao Feng}, {and}
  \bibinfo{person}{Jiancheng Lv}.} \bibinfo{year}{2023}\natexlab{}.
\newblock \showarticletitle{GPT-NAS: Neural Architecture Search with the
  Generative Pre-Trained Model}.
\newblock \bibinfo{journal}{\emph{arXiv preprint arXiv:2305.05351}}
  (\bibinfo{year}{2023}).
\newblock


\bibitem[Zhang et~al\mbox{.}(2023)]%
        {zhang2023automl}
\bibfield{author}{\bibinfo{person}{Shujian Zhang}, \bibinfo{person}{Chengyue
  Gong}, \bibinfo{person}{Lemeng Wu}, \bibinfo{person}{Xingchao Liu}, {and}
  \bibinfo{person}{Mingyuan Zhou}.} \bibinfo{year}{2023}\natexlab{}.
\newblock \showarticletitle{AutoML-GPT: Automatic Machine Learning with GPT}.
\newblock \bibinfo{journal}{\emph{arXiv preprint arXiv:2305.02499}}
  (\bibinfo{year}{2023}).
\newblock


\bibitem[Zhao et~al\mbox{.}(2023)]%
        {zhao2023survey}
\bibfield{author}{\bibinfo{person}{Wayne~Xin Zhao}, \bibinfo{person}{Kun Zhou},
  \bibinfo{person}{Junyi Li}, \bibinfo{person}{Tianyi Tang},
  \bibinfo{person}{Xiaolei Wang}, \bibinfo{person}{Yupeng Hou},
  \bibinfo{person}{Yingqian Min}, \bibinfo{person}{Beichen Zhang},
  \bibinfo{person}{Junjie Zhang}, \bibinfo{person}{Zican Dong},
  {et~al\mbox{.}}} \bibinfo{year}{2023}\natexlab{}.
\newblock \showarticletitle{A survey of large language models}.
\newblock \bibinfo{journal}{\emph{arXiv preprint arXiv:2303.18223}}
  (\bibinfo{year}{2023}).
\newblock


\bibitem[Zheng et~al\mbox{.}(2023)]%
        {zheng2023can}
\bibfield{author}{\bibinfo{person}{Mingkai Zheng}, \bibinfo{person}{Xiu Su},
  \bibinfo{person}{Shan You}, \bibinfo{person}{Fei Wang}, \bibinfo{person}{Chen
  Qian}, \bibinfo{person}{Chang Xu}, {and} \bibinfo{person}{Samuel Albanie}.}
  \bibinfo{year}{2023}\natexlab{}.
\newblock \showarticletitle{Can GPT-4 Perform Neural Architecture Search?}
\newblock \bibinfo{journal}{\emph{arXiv preprint arXiv:2304.10970}}
  (\bibinfo{year}{2023}).
\newblock


\bibitem[Zhou et~al\mbox{.}(2022)]%
        {zhou2022large}
\bibfield{author}{\bibinfo{person}{Yongchao Zhou}, \bibinfo{person}{Andrei~Ioan
  Muresanu}, \bibinfo{person}{Ziwen Han}, \bibinfo{person}{Keiran Paster},
  \bibinfo{person}{Silviu Pitis}, \bibinfo{person}{Harris Chan}, {and}
  \bibinfo{person}{Jimmy Ba}.} \bibinfo{year}{2022}\natexlab{}.
\newblock \showarticletitle{Large language models are human-level prompt
  engineers}.
\newblock \bibinfo{journal}{\emph{arXiv preprint arXiv:2211.01910}}
  (\bibinfo{year}{2022}).
\newblock


\end{thebibliography}

\end{document}